\documentclass[superscriptaddress,twoside,twocolumn,showpacs,preprintnumbers,amssymb,prl]{revtex4}
\usepackage{graphicx}
\usepackage{dcolumn}
\usepackage{bm}
\begin{document}
\title{Brittle Crack Roughness in Three-Dimensional Beam Lattices}
\author{Bj{\o}rn Skjetne}
 \affiliation{Department of Chemical Engineering,
              Norwegian University of Science and Technology,\\
              N-7491 Trondheim, Norway}
 \affiliation{Department of Physics,
              Norwegian University of Science and Technology,\\
              N-7491 Trondheim, Norway}
\author{Torbj{\o}rn Helle}
 \affiliation{Department of Chemical Engineering,
              Norwegian University of Science and Technology,\\
              N-7491 Trondheim, Norway}
\author{Alex Hansen}
 \affiliation{Department of Physics,
              Norwegian University of Science and Technology,\\
              N-7491 Trondheim, Norway}
\date{\today}
\begin{abstract}
The roughness exponent is reported in numerical simulations 
with a three dimensional elastic beam lattice. Two different
types of disorder have been used to generate the breaking 
thresholds, i.e., distributions with a tail towards
either strong or weak beams. Beyond the weak disorder regime
a universal exponent of $\zeta=0.59\pm0.01$ is obtained.
This is within the range $0.4\lesssim\zeta\lesssim0.6$ reported
experimentally for small scale quasi-static fracture, as
would be expected for media with a characteristic length scale. 
\end{abstract}
\maketitle
The interest in crack
morphology can be traced back to Mandelbrot et al., who
introduced a mathematical framework for describing rough 
surfaces in terms of fractal geometry~\cite{mand}. 
More specifically, 
crack surfaces have been shown to be self-affine in the sense
that they satisfy certain scaling laws, where a characteristic 
exponent governs the asymptotic behaviour~\cite{darc}.
Since the quantitative properties of crack surfaces have been
made readily accessible through the development of modern imaging 
techniques, predictions of computer models~\cite{smod} 
can be compared with the results of practical experiments to 
verify or discard theoretical assumptions.

Much of the theoretical work done so far has been based on the 
random fuse model~\cite{fuse}, 
i.e., a regular lattice of conducting elements which
irreversibly burn out once their individual thresholds have
been exceeded. Breakdown is driven by a potential
difference between two opposing boundaries and the analogy
of Kirchhoff's equations with linear elasticity is the
reason why this model is referred to as a scalar model of 
fracture. The result obtained for the roughness exponent in
simulations with the fuse model is
$\zeta=0.74(2)$ in two dimensions~\cite{hans}. Recently,
a somewhat different result, $\zeta=0.84(3)$, has 
been reported~\cite{zapp}.
Experimental studies in two dimensions have been carried out
by Poirer et al.~\cite{poir}, who considered a stacking of 
collapsible cylinders (drinking straws) to obtain $\zeta=0.73(7)$, 
by Kertesz et al.~\cite{kert} for tear lines in wet paper, 
obtaining $\zeta\approx0.73$, and by Eng{\o}y et al.~\cite{engo} 
for crack lines in thin wood plates, yielding a roughness
exponent of $\zeta=0.68(4)$. 
Results also differ according to the dimensionality, i.e.,
in three dimensions the fuse model result is 
$\zeta=0.62(5)$~\cite{batr}. This result, however, 
is very different from the universal value suggested by Bouchaud
et al.~\cite{bouc}, i.e., $\zeta=0.8$, a value which has been
confirmed in experimental studies~\cite{exp3}.
Hence the apparent agreement of the fuse model with experiment
in two dimensions is a coincidence. 

An important point
to be considered with the fuse model, however, is the
fact that it models electrical breakdown rather than brittle
elastic fracture.
A different model which incorporates elasticity is
the Born-model~\cite{born}. Here the material is modeled
as a network of elastic springs, each spring being free to 
rotate at its ends. In two dimensions $\zeta=0.64(5)$ is
obtained with this model~\cite{cald}.
As with the fuse model,
the exponent drops when going from two to three dimensions.
The result, $\zeta=0.5$~\cite{pari}, is also significantly 
different from the universal value of $\zeta=0.8$. Recent 
findings, however, indicate that there should exist two 
different regimes for the scaling behaviour of cracks. As 
was first shown by Milman et al.~\cite{milm}, a much
lower exponent $\zeta\approx0.5$ is found on small length
scales. Daguier et al.\ identified two self-affine fracture 
regimes, where $\zeta\sim0.5$ is
associated with small length scales and slow crack growth
and $\zeta=0.8$ involves dynamic effects and is associated
with larger length scales~\cite{velo}.

A problem with the Born model is that it is not
rotationally invariant~\cite{feng}. A different model which does not
suffer from this drawback is the elastic beam model~\cite{roux}.
Here the beams are rigidly connected at the nodes so as to 
preserve the angle between any two neighbouring beams. 
Rotations thus induce flexing and twisting deformations, 
while linear displacements induce transverse shear and axial 
forces. The exponent obtained with the beam model is 
$\zeta=0.86(3)$ in two dimensions~\cite{skje}, 
and shows that the scalar and vectorial
descriptions of fracture need not necessarily produce 
similar results. Since the beam model provides a more realistic
description of brittle elastic fracture 
than does either the fuse model or the Born model 
it is of great interest to
see what the result is in three dimensions.

\begin{figure}
\centering
\includegraphics[scale=0.345]{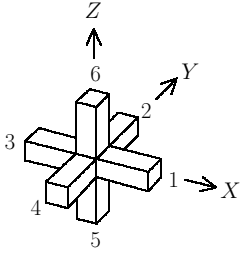}
\caption{Enumeration scheme on the beams of a cube lattice
         connecting node~$i$ to its $j=1$--$6$ nearest neighbours,
         showing the coordinate system with $i$ as its origin.
         \label{6beam}}
\end{figure}
The lattice used in our calculations
is a regular cube of size $L\times L\times L$, where
each node is connected to its nearest neighbours
by linearly elastic beams of unit length. 
Forces acting on the nodes have been deduced from the effect 
a concentrated end-load has on a beam with no 
end-restraints~\cite{roar}. A coordinate system is placed 
on each node, and the enumeration of the connecting beams 
follows an anti-clockwise 
scheme within the $XY$-plane, i.e., beginning with the beam which 
lies along the positive $X$-axis and ending with that
which extends upwards along the positive $Z$-axis, 
see Fig.~\ref{6beam}.

At each stage of the breaking process, the updated displacements
for each node is obtained from
\begin{eqnarray}
    \sum_{j}{\bf D}_{ij}
             \left[\begin{array}{l}
                      x_{i}\\
                      y_{i}\\
                      z_{i}\\
                      u_{i}\\
                      v_{i}\\
                      w_{i}\\
                    \end{array}
             \right]=\lambda
             \left[\begin{array}{l}
                      X_{i}\\
                      Y_{i}\\
                      Z_{i}\\
                      U_{i}\\
                      V_{i}\\
                      W_{i}\\
                   \end{array}
             \right],
              \label{ma3x}
\end{eqnarray}
which is solved iteratively via relaxation using the conjugate 
gradient method. 
This minimizes the 
elastic energy to obtain those displacements for which the
sum of forces and moments on each node vanish, i.e. the
mechanical equilibrium.

Six terms contribute to each of the force components
in Eq.~(\ref{ma3x}). With the neighbouring nodes fixed, a
translational displacement $\delta x=x_{j}-x_{i}$ of node~$i$ 
induces axial forces in beams 1 and 3, and transverse forces in
beams 2, 4, 5 and 6. With the expressions for force and moment 
defined in Refs.~\cite{skje} and~\cite{herr}, this gives
\begin{eqnarray}
    X_{i}={_{x}}A_{i}^{(1)}+{_{x}}A_{i}^{(3)}
         +\sum_{j\ne1,3}{_{x}}S_{i}^{(j)},
           \label{xcomp}
\end{eqnarray}
with similar expressions for $Y_{i}$ and $Z_{i}$.
Likewise, a rotational displacement $\delta u=u_{j}-u_{i}$ 
about the $X$-axis causes a torsional moment in
beams 1 and 3, and flexure in beams 2, 4, 5 and 6. Hence,
\begin{eqnarray}
    U_{i}={_{x}}T_{i}^{(1)}+{_{x}}T_{i}^{(3)}
         +\sum_{j\ne1,3}{_{x}}M_{i}^{(j)},
           \label{ucomp}
\end{eqnarray}
now with similar expressions for $V_{i}$ and $W_{i}$.

To express the thirty-six force components in Eq.~(\ref{ma3x}) more
compactly,
\begin{eqnarray}
    r_{j}=\prod_{n=0}^{j-1}\bigl(-1\bigr)^{n}
\end{eqnarray}
and
\begin{eqnarray}
    s_{j}=\bigl(-1\bigr)^{j}r_{j}
\end{eqnarray}
are defined for notational convenience, to keep track of the 
signs. The Kronecker delta, moreover, has been used 
to construct
\begin{eqnarray}
    \widehat{\lambda}_{s,t}^{(j)}=
     \delta_{sj}+\delta_{tj}
      \label{inc}
\end{eqnarray}
and
\begin{eqnarray}
    \widehat{\chi}_{s,t}^{(j)}=
     \delta_{kj}
      \bigl(1-\delta_{sj}\bigr)\bigl(1-\delta_{tj}\bigr),
       \label{exc}
\end{eqnarray}
operators which include or exclude $s$ and $t$ from the sum
over neighbouring beams. 

Eq.~(\ref{xcomp}) can then be stated on the form
\begin{widetext}
\begin{eqnarray}
    X_{i}=\sum_{j=1}^{6}
           \frac{1}{\alpha}
            \widehat{\lambda}_{1,3}^{(j)}\delta x
         +\sum_{j=1}^{6}
           \frac{1}
                {\beta+\frac{\gamma}{12}}\widehat{\chi}_{1,3}^{(j)}
             \biggl\{\delta x
             -\frac{r_{j}}{2}
         \Bigl[\widehat{\lambda}_{5,6}^{(j)}\bigl(v_{i}+v_{j}\bigr)+
               \widehat{\lambda}_{2,4}^{(j)}\bigl(w_{i}+w_{j}\bigr)
         \Bigr]
             \biggr\},
\end{eqnarray}
where, exchanging 
indices (1,3) and (2,4), and interchanging $v$ with $u$ and 
$x$ with $y$, $Y_{i}$ is obtained from $X_{i}$. 
In the same way, $Z_{i}$ is obtained from $X_{i}$ by an exchange of the
indices (1,3) and (5,6), by interchanging $w$ with $u$ and
$x$ with $z$, and by replacing 
$r_{j}$ with $-s_{j}$. 

Next, Eq.~(\ref{ucomp}) 
can be stated on a similar form, i.e.,
\begin{eqnarray}
    U_{i}=\sum_{j=1}^{6}
          \frac{1}{\rho}
          \widehat{\lambda}_{1,3}\delta u
         +\sum_{j=1}^{6}
          \frac{1}{\beta+\frac{\gamma}{12}}
          \widehat{\chi}_{1,3}
          \biggl\{
             \frac{\beta}{\gamma}\delta u
            +\frac{r_{j}}{2}
             \Bigl[\hspace{0.5mm}
                 \widehat{\lambda}_{5,6}\delta y
                -\widehat{\lambda}_{2,4}\delta z
             \Bigr]
            -\frac{1}{3}\bigl(u_{i}+\frac{1}{2}u_{j}\bigr)
          \biggr\},
\end{eqnarray}
\end{widetext}
where $V_{i}$, for angular displacements about the
$Y$-axis, is obtained from $U_{i}$ by the same 
substitutions as in $X_{i}\rightarrow Y_{i}$.
Likewise, the expression for $W_{i}$, 
relevant to angular displacements about the $Z$-axis, follows
from $U_{i}$ by use of the $X_{i}\rightarrow Z_{i}$ 
substitutions, now however, without changing $r_{j}$ to $-s_{j}$. 
Instead, the change from $z$ to $x$ is made so that
$-\delta x$ replaces $\delta z$. 
\begin{figure} [b]
\centering
\includegraphics[angle=-90,scale=0.19]{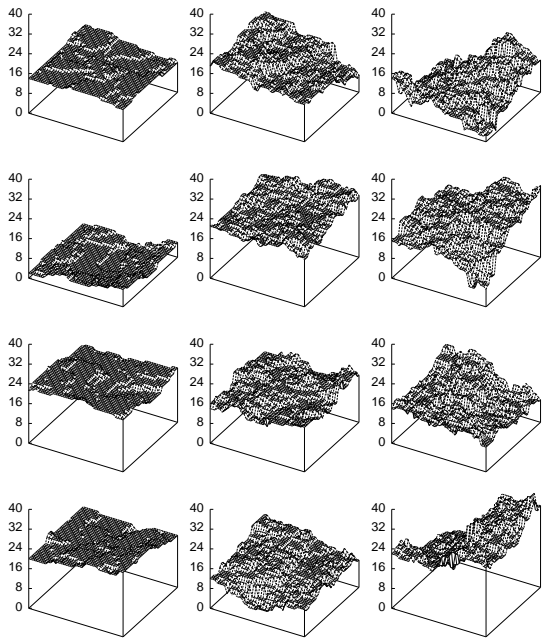}
\caption{Fracture surfaces, showing
         the lower remaining part of a cube lattice of size
         $L=40$ after it has broken completely. Four samples are 
         included for $D=1$ on the left, for $D=2$
         at center, and for $D=-4$ on the right.
         \label{4cube}}
\end{figure}

Breaking thresholds are generating by raising a random 
number $r$ on the interval $[0,1]$ to a power $D$. The 
magnitude of $D$ then corresponds to the strength of disorder,
with the distribution having a tail towards weak beams
for $D>0$ and towards strong beams for $D<0$.
This is the most simple way of incorporating scale invariance,
i.e., those mechanisms responsible for the
multifractal behaviour of disordered systems,
into the distribution of thresholds~\cite{dxdx}. For each
sample two casts are generated for the thresholds, one
for the flexural strength and one for the axial strength.

\begin{figure}
\centering
\includegraphics[angle=0,scale=0.485]{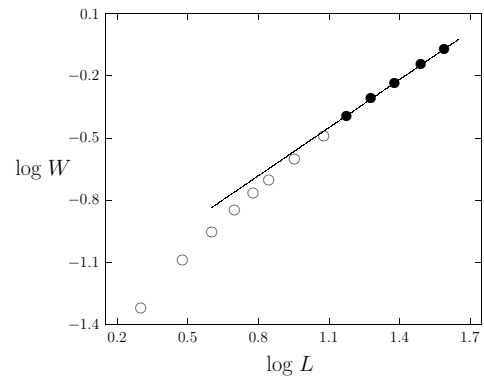}
\caption{Logarithmic plot, showing the roughness, $W$, as a 
         function of the system size, $L$, for disorder
         $D=1$. Filled circles denote those points to which 
         the straight line, with slope $\zeta=0.77(1)$, 
         has been fit.
         \label{dx01}}
\end{figure}
\begin{figure} [b]
\centering
\includegraphics[angle=0,scale=0.485]{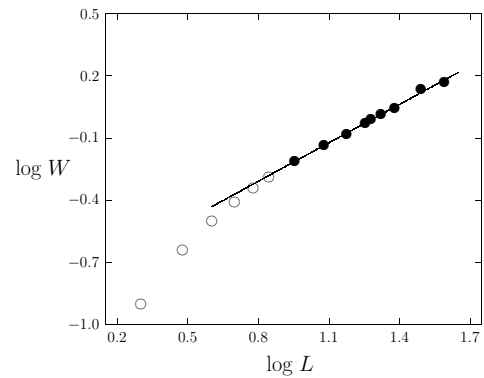}
\caption{Logarithmic plot, showing the roughness, $W$, as a 
         function of the system size, $L$, for disorder
         $D=2$. Filled circles denote those points to which 
         the straight line, with slope $\zeta=0.62(2)$, 
         has been fit.
         \label{dx02}}
\end{figure}
Fracture is initiated by imposing a uniform displacement 
on all nodes defining the top surface of the cube. The first
beam to break is the one with the weakest axial strength,
whereupon the location of subsequent breaks depends on 
a complex
interplay between quenched disorder and the constantly
evolving non-uniform stress-field. 
Each time a beam is
removed from the lattice, Eq.~(\ref{ma3x}) is used to
obtain the new equilibrium displacements. This is
continued until a surface appears which divides the sample
in two separate parts, see Fig.~\ref{4cube}.

\begin{figure} [t]
\centering
\includegraphics[angle=0,scale=0.485]{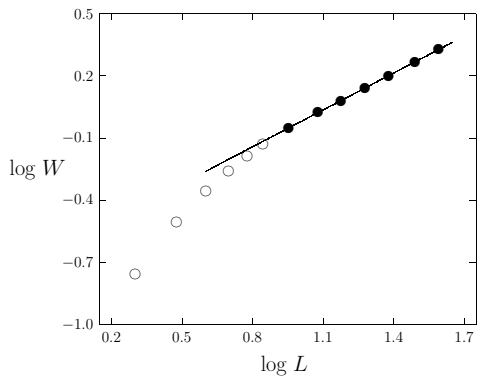}
\caption{Logarithmic plot, showing the roughness, $W$, as a 
         function of the system size, $L$, for disorder
         $D=4$. Filled circles denote those points to which 
         the straight line, with slope $\zeta=0.59(1)$, 
         has been fit.
         \label{dx04}}
\end{figure}

Results obtained for $D=1$, $2$ and $4$ are shown in 
Figs.~\ref{dx01},~\ref{dx02} and~\ref{dx04}, respectively, 
where the roughness $W$ has been calculated for systems ranging 
in size from $L=3$ to $L=40$. Disregarding finite-size
effects for small $L$, the relationship between $L$ and 
$W$ is clearly self-affine. 
Hence, defining the roughness as the root-mean-square of
the variance perpendicular to the (average) fracture
plane, i.e.,
\begin{eqnarray}
    W_{x}(L)=\left\langle\frac{1}{L}\sum_{i=1}^{L}z_{x}(i)^{2}-
          \biggl[\frac{1}{L}\sum_{i=1}^{L}z_{x}(i)\biggr]^{2}
           \right\rangle^{1/2},
\end{eqnarray}
we have $W\sim L^{\zeta}$, with the results of $W_{x}$ and 
$W_{y}$ being statistically the same.

The exponent in Fig.~\ref{dx01} is $\zeta=0.77(1)$, and does 
not belong to the self-affine regime of fracture. This
is because $D=1$ in a three dimensional lattice generates 
too little disorder. The qualitative difference between the 
fracture surfaces obtained for $D=1$ and higher values of $D$ 
is shown in Fig.~\ref{4cube}. 
A universal exponent emerges as the disorder is
increased, however, with the exponents obtained for $D=2$ and
$D=4$ being $\zeta=0.62(2)$ and $\zeta=0.592(5)$, respectively.

\begin{figure} [b]
\centering
\includegraphics[angle=0,scale=0.485]{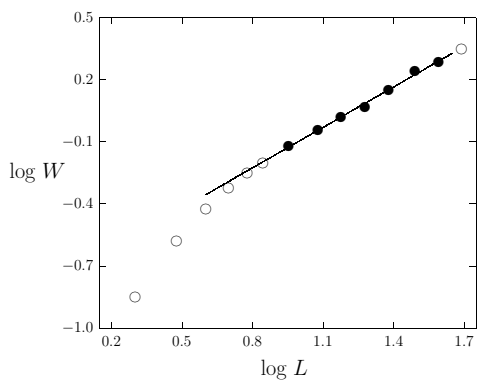}
\caption{Logarithmic plot, showing the roughness, $W$, as a 
         function of the system size, $L$, for disorder
         $D=-4$. Filled circles denote those points to which 
         the straight line, with slope $\zeta=0.65(2)$, 
         has been fit.
         \label{dy04}}
\end{figure}

The result also seems to hold for disorders with a tail
towards strong beams. For $D=-4$ the exponent 
is $\zeta=0.65(2)$, see Fig.~\ref{dy04}. 
Although this is slightly different
from the best value obtained for $D>0$, which we take to 
be $\zeta=0.59(1)$, the small discrepancy is probably due to 
the fact that $D=-4$ lies just within the transient regime 
towards strong disorders for distributions of type
$D<0$. In two dimensions the $D<0$ stress-strain curve displays
a cross-over towards stable crack-growth when 
$|D|\sim2$. 
This cross-over value is probably higher in the cube lattice,
due to the stronger constraints imposed on the 
crack path in three dimensions. 
 
To summarize, the roughness exponent $\zeta=0.59(1)$ is
obtained numerically for a cubic lattice of elastic beams. 
This is lower than the experimental value relevant to 
large scales, $\zeta=0.8$, but within the range 
reported for small scales, $0.4<\zeta<0.6$. This lends
further evidence to the existence of a different scaling
regime for slow crack growth involving 
characteristic, or ``small'', length
scales~\cite{rela}. The agreement of our result 
with the exponents reported in this regime stems from 
the discretization in terms of 
a beam lattice, i.e., a system which involves a characteristic 
intermediate length scale, as in the case of Cosserat
elasticity.


\begin{thebibliography}{99}
\bibitem{rela} The term small here is relative -- 
               for instance, $\zeta\sim0.5$
               has also been reported for cracks in sandstone
               where the typical grain-size is large compared with the 
               length scales usually associated with this
               exponent. See, e.g.,
               J. M. Boffa, C. Allain, and J. P. Hulin,
               Eur.\ Phys.\ J -- Appl.\ Phys.\ {\bf 2}, 281 (1998).
\end{thebibliography}
\end{document}